\documentclass[12pt,apj]{emulateapj}

\usepackage{subfigure}

\begin{document}

\shorttitle{Internetwork chromospheric bright grains observed with IRIS}
\shortauthors{Mart\'inez-Sykora et al.}
\title{Internetwork Chromospheric Bright Grains Observed With IRIS and SST}

\author{Juan Mart\'inez-Sykora$^{1,2}$}
\email{j.m.sykora@astro.uio.no}
\author{Luc Rouppe van der Voort$^{3}$}
\author{Mats Carlsson$^{3}$}
\author{Bart De Pontieu$^{1,3}$}
\author{Tiago M. D. Pereira$^{3}$}
\author{Paul Boerner$^{1}$}
\author{Neal Hurlburt$^{1}$}
\author{Lucia Kleint$^{4}$}
\author{James Lemen$^{1}$}
\author{Ted D. Tarbell$^{1}$}
\author{Alan Title$^{1}$}
\author{Jean-Pierre Wuelser$^{1}$}
\author{Viggo H. Hansteen$^{3,1}$}
\author{Leon Golub$^{5}$}
\author{Sean McKillop$^{5}$}
\author{Kathy K. Reeves$^{5}$}
\author{Steven Saar$^{5}$}
\author{Paola Testa$^{5}$}
\author{Hui Tian$^{5}$}
\author{Sarah Jaeggli$^{6}$}
\author{Charles Kankelborg$^{6}$}

\affil{$^1$ Lockheed Martin Solar and Astrophysics Laboratory, Palo Alto, CA 94304}
\affil{$^2$ Bay Area Environmental Research Institute, Sonoma, CA}
\affil{$^3$ Institute of Theoretical Astrophysics, University of Oslo, P.O. Box 1029 Blindern, N-0315 Oslo, Norway}
\affil{$^4$ University of Applied Sciences and Arts Northwestern Switzerland, 5210 Windisch, Switzerland}
\affil{$^5$ Harvard-Smithsonian Center for Astrophysics, 60 Garden Street, Cambridge, MA, 02138}
\affil{$^6$ Department of Physics, Montana State University, Bozeman, P.O. Box 173840, Bozeman, MT 59717}

\newcommand{\myemail}{juanms@astro.uio.no}
\newcommand{\komment}[1]{\texttt{#1}}
\newcommand{\newch}[1]{{#1}}
\def\Halpha{\mbox{H\hspace{0.1ex}$\alpha$}}
\def\caia{\mbox{\ion{Ca}{2}~8542~\AA}}
\def\caka{\mbox{\ion{Ca}{2}~K~3933~\AA}}
\def\caha{\mbox{\ion{Ca}{2}~H~3968~\AA}}
\def\mgka{\mbox{\ion{Mg}{2}~k~2796~\AA}}
\def\mgha{\mbox{\ion{Mg}{2}~h~2803~\AA}}
\def\cai{\mbox{\ion{Ca}{2}~8542~\AA}}
\def\cak{\mbox{\ion{Ca}{2}~K}}
\def\cah{\mbox{\ion{Ca}{2}~H}}
\def\mgk{\mbox{\ion{Mg}{2}~k}}
\def\mgh{\mbox{\ion{Mg}{2}~h}}
 \def\fe1{\mbox{\ion{Fe}{1}~6302~\AA}}
\def\si4{\mbox{\ion{Si}{4}}}
\def\sia4{\mbox{\ion{Si}{4}~1403~\AA}}

\begin{abstract}

  The Interface Region Imaging Spectrograph (IRIS) reveals small-scale
  rapid brightenings in the form of bright grains all over
  coronal holes and the quiet sun. These bright grains are seen with the
  IRIS 1330~\AA, 1400~\AA\ and 2796~\AA\ slit-jaw filters. We combine
  coordinated observations with IRIS and from the ground with the
  Swedish 1-m Solar Telescope (SST) which allows us to have
  chromospheric (\caia, \caha, \Halpha, and \mgka), and transition
  region (\ion{C}{2}~1334~\AA, \sia4) spectral imaging, and
  single-wavelength Stokes maps in \fe1 at high spatial ($0\farcs33$), 
  temporal and spectral resolution. We conclude that the IRIS slit-jaw grains are
  the counterpart of so-called acoustic grains, i.e., resulting from
  chromospheric acoustic waves in a non-magnetic environment. We
  compare slit-jaw images with spectra from the IRIS spectrograph. We
  conclude that the grain intensity in the 2796~\AA\ slit-jaw filter
  comes from both the \mgk\ core and wings. The signal in the
  \ion{C}{2} and \ion{Si}{4} lines is too weak to explain the presence
  of grains in the 1300 and 1400~\AA\ slit-jaw images and we conclude
  that the grain signal in these passbands comes mostly from the
  continuum. Even though weak, the characteristic shock \newch{signatures} of
  acoustic grains can often be detected in IRIS \ion{C}{2}
  spectra. For some grains, spectral signature can be found in IRIS
  \si4. This suggests that upward propagating acoustic waves sometimes
  reach all the way up to the transition region.

\end{abstract}


\section{Introduction}

Waves can be observed throughout all layers of the solar atmosphere. 
Understanding the 
properties of waves is important because of their impact on chromospheric 
and coronal heating and the solar wind. 
In addition, waves have the potential to serve as a diagnostic to
measure magneto-thermal properties and to decide which physical
processes are important in the different layers in the solar
atmosphere. 

In this study, we concentrate on the internetwork of the quiet Sun,
where observations in chromospheric spectral lines such as \caha\ and
\caka\ are dominated by small grains: short-lived (100~s or less), 
sub-arcsecond regions with enhanced intensity \citep[see e.g., the
extensive review by][]{Rutten:1991fk}. \citet{Rutten:1999xy,Handy:1999nr,Tian:2010qf}, among others, 
show that the TRACE near-UV also reveals nicely the bright grains. 
These grains result from acoustic waves that propagate upward in the
non-magnetic environment and turn into shocks as they reach the chromosphere
\citep[e.g.,][]{Carlsson:1992kl,Carlsson+Stein1994,Carlsson:1997tg}. 
\cak\ spectral profiles emerging from acoustic shocks, so called K-grains, were studied 
under non-LTE conditions in 1D hydrodynamic models by 
\citet{Carlsson:1992kl,Carlsson:1997tg} and in 3D under LTE conditions
by \citet{Wedemeyer:2004}. 

Propagating from the upper layers of the photosphere, the acoustic waves
steepen into acoustic shocks in the chromosphere 
due to the steep decrease in density \citep[][among
others]{Carlsson:1992kl,Carlsson:1997tg,Wedemeyer:2004,Danilovic:2014fk}.  

One may expect that these grains could be traced throughout the chromosphere 
and transition region. Assuming that the shocks continue to travel upward, 
the density decrease through the chromosphere and transition region would 
lead to sharpening of the shock and an increase in shock amplitude. 
However, observations from the Solar Ultraviolet Measurements 
of Emitted Radiation \citep[SUMER,][]{Wilhelm:1995fk} on board of the 
Solar and Heliospheric Observatory \citep[SOHO,][]{Domingo:1995sf},
and from the Vacuum Tower Telescope \citep[VTT,][]{Bendlin:1995rm} 
indicated that the shocks that produced 
K-grains apparently barely have a transition region counterpart 
\citep{Steffens:1997ys,Carlsson:1997ys,Judge:1997zr}. 
\citet{Carlsson:1997ys} could trace acoustic wave related
oscillations in UV continua and spectral lines of neutral species
but only occasionally and faintly in spectral lines of singly
ionized species (\ion{C}{2}~1335~\AA). 
Spectral lines from double ionized species, the ``hottest" lines in
their sample, did not show signs of acoustic wave related oscillations 
in the internetwork.
\citet{Judge:1997zr} concluded that upward-propagating acoustic shock
waves do not contribute significantly to the heating of the lower
transition region. 
The question then remains how these acoustic waves are dissipated
before they reach the transition region and to what extent they
contribute to chromospheric heating in the quiet Sun. 

In this paper we focus on acoustic chromospheric waves in quiet sun
and coronal hole internetwork using some of the highest resolution
observations to date: in the UV from space with the Interface Region Imaging 
Spectrograph \citep[IRIS,][]{De-Pontieu:2014vn} and the strong
chromospheric lines \Halpha\ and \caia\ from the ground with the Swedish 1-m
Solar Telescope \citep[SST,][]{Scharmer:2003ve}.
These instruments provide a combination of high spatial, spectral and
temporal resolution.  

The paper layout is as follows, the data processing of the various instruments 
is described in Section~\ref{sec:data}. In Section~\ref{sec:obsres} we follow up with 
the observational results which focus on 1) the properties of 
bright chromospheric internetwork grains (Section~\ref{sec:prop}); 2)
the origin of the 
emission of these grains (Section~\ref{sec:emiss}); and 3) the spectral analysis 
(Section~\ref{sec:dopp}). Finally, we discuss the results and conclusions 
in Section~\ref{sec:dis}. 

\section{Observations}
\label{sec:data}

IRIS obtains spectra in passbands from 1332--1358~\AA\ \newch{(with spectral 
pixel size of 12.98~m\AA), 1389--1407~\AA\ (with spectral pixel size 
of 12.72~m\AA) and 2783--2834~\AA\ (with spectral pixel size of 
25.46~m\AA)} including bright spectral lines formed in the chromosphere,  
e.g., \mgha\ and k~2796~\AA,
in the upper chromosphere/lower transition region (\ion{C}{2} 1334/1335~\AA) 
and in the transition region (e.g., \ion{Si}{4}~1394/1403~\AA). Spectral rasters sample regions up 
to 130\arcsec$\times$175\arcsec\ at a variety of spatial samplings
(from $0\farcs33$ and up). 
In addition, IRIS can take slit-jaw images (SJI) with different filters that
have spectral windows dominated by emission from these spectral lines. 
SJI 2796 is centered on \mgk\ at 2796~\AA\ and has a 4~\AA\ bandpass, 
SJI 1330 is centered at 1340~\AA\ and has a 55~\AA\ bandpass, and
SJI 1400 is centered at 1390~\AA\ and has a 55~\AA\ bandpass.
For more information on IRIS, we refer the reader to 
\citet{De-Pontieu:2014vn}. We analyze both sit-and-stare\newch{, i.e, with the slit 
pointing at a fixed solar region and continuously tracking solar rotation,} and 
large spatial raster IRIS observations.

The sit-and-stare observations were obtained on 2013 September 
22 from 07:34:30 to 11:04:13 UT where the IRIS slit was kept at a fixed
location to maximize the temporal cadence. The cadence of the 
spectral observations was 5~s with an exposure time of 4~s. Slit-jaw 
images with the 1330~\AA, 1400~\AA\ and 2796~\AA\ filters 
were taken every 10~s. Calibrated level 2 data was used in our study, 
i.e., dark subtraction, flat field correction, and geometrical correction 
have been taken into account \citep{De-Pontieu:2014vn}. The target 
was a coronal hole at $(x, y) \sim (537\arcsec,282\arcsec)$. 
For this dataset, we acquired coordinated observations with the 
Swedish 1-m Solar Telescope \citep[SST,][]{Scharmer:2003ve} 
on La Palma using the CRisp Imaging SpectroPolarimeter 
\citep[CRISP,][]{Scharmer:2008zv} and imaging in the blue beam. 
CRISP provides us with spectrally resolved imaging in \newch{\Halpha\ and
\caia,} and single-wavelength Stokes $I,Q,U,V$ images in the wing 
($-48$~m\AA) of \fe1.   
The time range for CRISP was 08:09:00--10:10:47 UT.
For the period with best seeing (08:20:39 -- 09:08:43) we acquired
upper photospheric/chromospheric images in a \cah\ filter (1.1~\AA\ FWHM)
centered on the line core and pure photospheric images from a
wide-band filter (10~\AA\ FWHM) centered on 3953.7~\AA, the bump 
between the Ca K and H cores. The \caia\ spectral line was sampled at 25 spectral 
positions within $\pm1200$~m\AA\ \newch{of the core} with a sampling of 100~m\AA. The \Halpha\ 
line was sampled at 15 spectral positions within $\pm1400$~m\AA\ with a sampling 
of 200~m\AA. \newch{A sharp and stable \newch{time-series} was achieved with aid from adaptive
optics, image restoration using the Multi-Object Multi-Frame Blind 
Deconvolution method \citep{van-Noort:2005uq}, and time series
processing of the reconstructed images that includes de-rotation,
rigid  alignment and de-stretching.
For the data processing we follow the CRISP reduction pipeline
\citep{de-la-Cruz-Rodriguez:2014pv} which includes procedures described by
\citet{2013A&A...556A.115D}, \citet{2012A&A...548A.114H}, 
\citet{2008A&A...489..429V}, and \citet{1994ApJ...430..413S}.}

We aligned the SST data to the IRIS observations by scaling down to
IRIS image scale (0\farcs16) and through cross-correlation of SST
\cai\ wing  \newch{($\Delta \lambda=-1$~\AA)} and IRIS \mgk\ SJI 2796 images. The accuracy of the 
alignment was found to be on the level of the IRIS pixel size.
The overlapping field of view between IRIS and SST was roughly 
25\arcsec$\times$35\arcsec. 

We also make use of very large dense IRIS level 2 raster data with 30~s exposures 
taken on 2014 February 25 at 20:50:31 UT and on 2013 
October 22 at 11:30:30 UT. Each of the rasters has 400 steps and scans 
$132\arcsec\times 175\arcsec$ in three and a half hours with a spatial pixel-size
of $0\farcs16$ along the slit and $0\farcs35$ as raster step size. Both rasters were
on the quiet sun, the first one centered at ($x,y$) = (77\arcsec,$-$72\arcsec) and the 
second one at ($x,y$) = ($-$304\arcsec,$-$109\arcsec). To allow for a 
reasonable 32~s cadence for the slit jaw images, only two SJI filters were 
used in each raster: the 1330 and 1400~\AA\ SJIs were used in the first raster, 
and the 1400 and 2796~\AA\ SJIs used in the second raster. We analyze both 
rasters here in order to have observations in all three filters.

Solar Dynamic Observatory/Helioseismic and Magnetic Imager 
\citep[SDO/HMI][]{Scherrer:2012qf} data has been used for context of the same region and time 
such as the two deep exposure IRIS raster observations (2014 February 25 at 20:50:30 UT, 
and 2013 October 22 at 11:30:30 UT). We use the magnetic field along the line of sight 
distributed with the \texttt{ssw\_jsoc\_time2data.pro } and \texttt{read\_sdo.pro} and the data 
was processed using \texttt{aia\_prep.pro}. 

\section{Results}~\label{sec:obsres}

\begin{figure*}
  \includegraphics[width=0.99\textwidth]{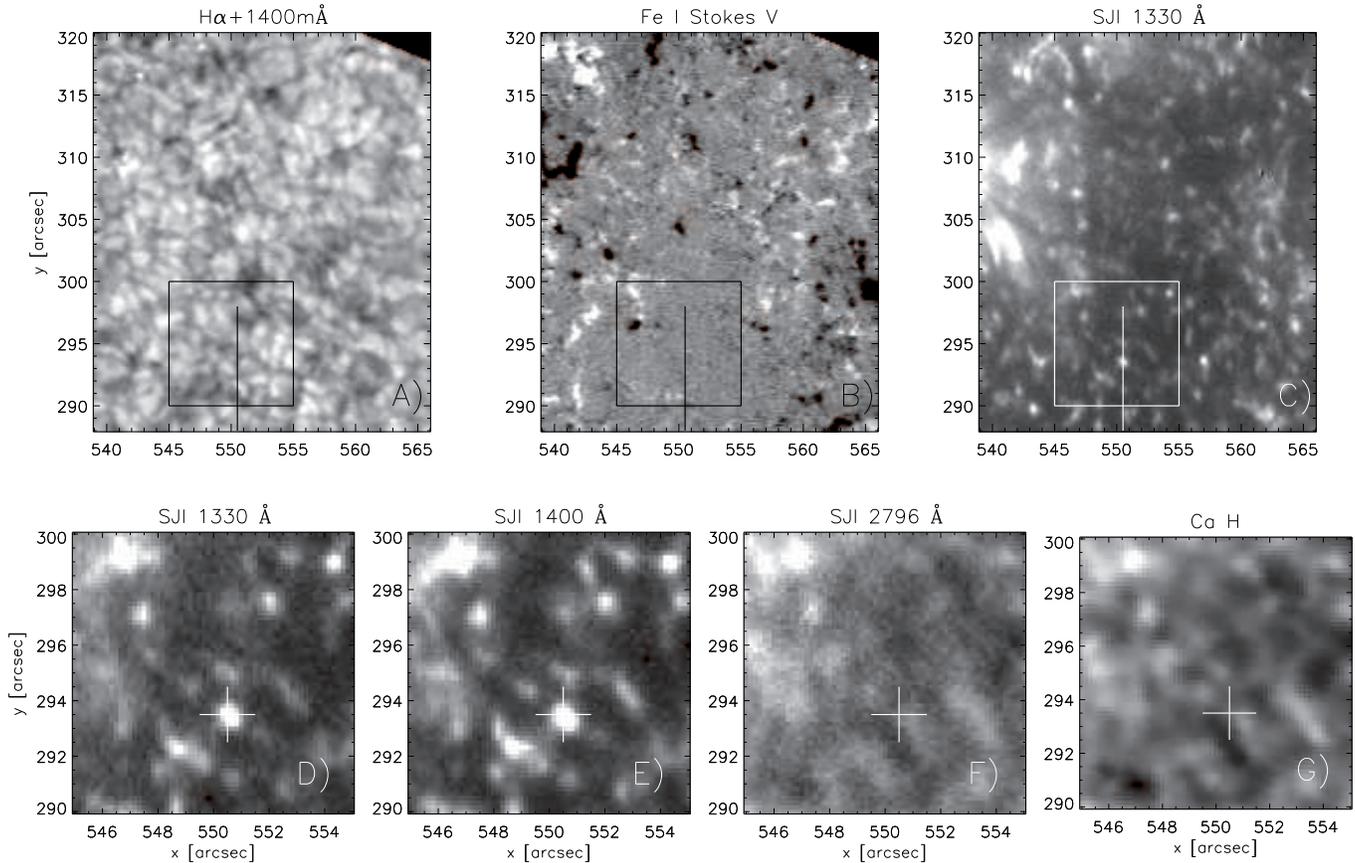}
 \caption{\label{fig:sjixy} Chromospheric bright grains (CBGs) in an
   internetwork region taken the 2013 September 22 at \newch{08:28:47}UT 
   \newch{(which corresponds to $t=3250$~s in Figures~\ref{fig:sjixt} and~\ref{fig:sji1d})}
   centered at $(x,y)=(537\arcsec,282\arcsec)$. 
   The observed region shown in the top panels
   is very quiet as can be inferred from the photospheric \Halpha\ far
   wing image at +1400~m\AA\ (panel A), and the \fe1\ Stokes V
   magnetogram (panel B). The IRIS SJI 1330~\AA\ image (panel C) is
   dominated by small bright grains scattered over the FOV. The
   vertical line in the top panels marks the location of the space/time diagram
   in Figure~\ref{fig:sjixt}. The rectangular box marks a region
   with particular low polarization signal that is shown at larger magnification in the panels in the
   bottom row. CBGs in IRIS SJI 1330~\AA\ (panel D), SJI 1400~\AA\ 
   (panel E), SJI 2796~\AA\ (panel F) and \cah\ (panel G) core maps
   overlap in time and space. The CBGs have a typical size of
   1\arcsec\ and lifetime of 1.5~min (see the corresponding Movie~1). The crosses in the bottom panels
   mark the location used for which the light curves are shown in Figure~\ref{fig:sji1d}.}  
\end{figure*}

Small-scale and dynamic bright grains dominate images and time series
of quiet Sun and coronal holes observed with the IRIS 1400~\AA,
1330~\AA, and 2796~\AA\ SJI filters (e.g., see
Figures~\ref{fig:sjixy},~\ref{fig:sjixt}, and~\ref{fig:shmi}). 
The properties of these chromospheric bright grains (CBGs), 
the origin of their emission in these filters, and spectral analysis are described 
in the following sections \ref{sec:prop}--\ref{sec:dopp},
respectively. 

\subsection{Properties of the CBGs}~\label{sec:prop}

\begin{figure*}
  \includegraphics[width=0.95\textwidth]{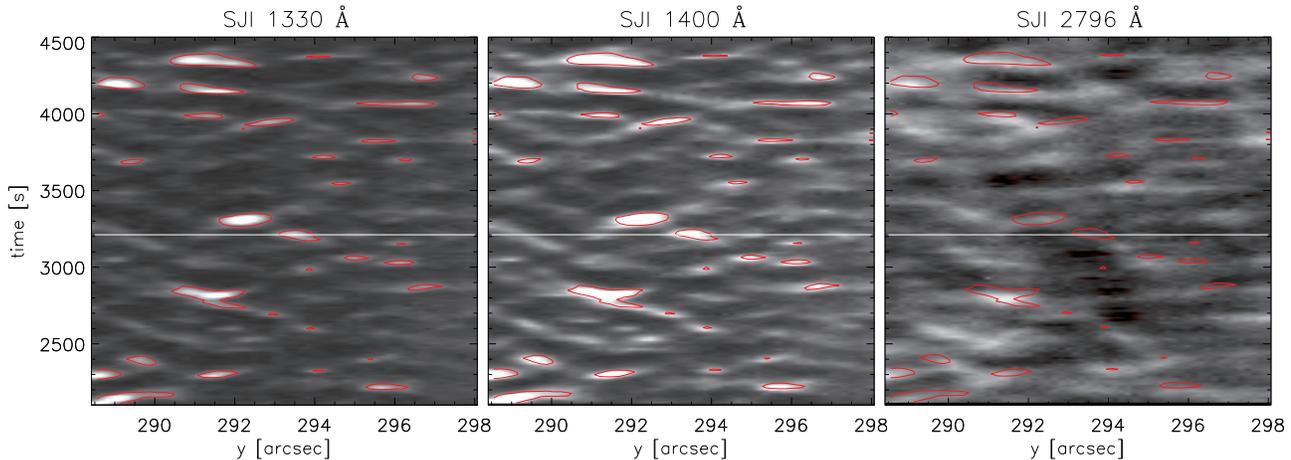}
 \caption{\label{fig:sjixt} Internetwork space/time diagram of IRIS
   SJI 1330~\AA\ (left), 1400~\AA\ (middle) and 2796~\AA\ (right).  The 
   red contours correspond to SJI 1330~\AA . The selected region is marked 
   in Figure~\ref{fig:sjixy}. \newch{The white line marks the time for which images are shown in Figure~\ref{fig:sjixy}}.}
\end{figure*}

We use the CRISP \fe1\ Stokes V maps to select the CBGs
observed with the IRIS SJI filters that are located in internetwork regions
with weak magnetic fields and to make sure 
that they are not related to isolated
magnetic bright points or network regions (see panel B in Figure~\ref{fig:sjixy}). 
Therefore, we do not study the isolated magnetic bright points studied by 
\citet{Sivaraman:1982sf,Sivaraman:2000rm,de-Wijn:2008hl}.
The typical size of the CBGs is of the order of $0\farcs5-1\farcs5$ with, in general,  
a roundish shape. Others are elongated such as the example located at  
$(x,y)=(553\arcsec,294\arcsec)$ (see panels D--G). 
The elongation can get as long as intergranular lanes ($\sim2$\arcsec). 
CBGs observed with SJI 1330~\AA\ 
(panel D), and 1400~\AA\ (panel E) are clearly more roundish than those 
observed  with SJI 2796~\AA\ (panel F) and in \cah\ (panel G). 

\begin{figure}
  \includegraphics[width=0.49\textwidth]{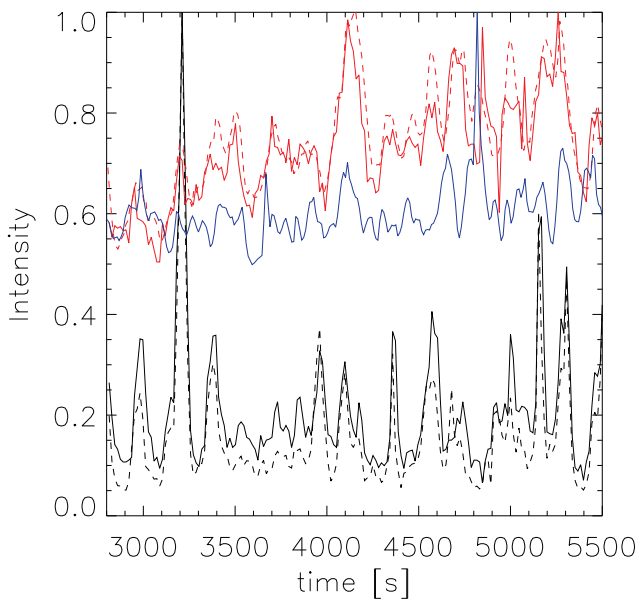}
 \caption{\label{fig:sji1d} Light curves for a selected position in an
   internetwork region with very low magnetic activity
   (marked with the cross in the bottom panels in
   Figure~\ref{fig:sjixy}): SJI 1330~\AA\ (solid black line), 1400~\AA\ (dashed black line), 
   2796~\AA\ (solid red line) filters and \cah\ (dashed red line) and integrated
   line core of \cai\ (solid blue line).  All light curves are normalized.}
\end{figure}

The CBGs observed in each SJI channel and \cah\ correspond to the same
feature, this is illustrated in Figure~\ref{fig:sjixy} and in movie~1. 
\newch{Overlap of the CGBs can also be discerned in space and time in all 
three channels (see Figure~\ref{fig:sjixt})}. In addition, they show a nice correlation with the
\cai\ line core intensity. 
The structures are highly similar between the 
SJI 1330 and 1400~\AA\ filters and between SJI 2796~\AA\ and \cah\ 
(see Figure~\ref{fig:sjixy} and Movie~1). 
Between these two sets of images (one set IRIS FUV SJIs and the
other set the longer wavelength passbands, i.e., SJI~2796~\AA\ and \cah), 
the brightenings seem to come from 
the same feature, but they show some differences in structure and brightness. 
Therefore, the CBG emission in the SJI 1330 and 1400~\AA\ filters 
is probably from a different atmospheric 
region than  SJI 2796~\AA\ and \cah, but still from the same features 
(see Section~\ref{sec:emiss}). We note that there seems to be a trend 
in the time evolution of the CBG's emissions (Figure~\ref{fig:sjixt} 
and~\ref{fig:sji1d}) being most of the time first seen in SJI 2796~\AA\  
and with longer duration than in the FUV channels. 
Within the context of upward propagating waves, this can be
interpreted as the signal of the SJI 2796 filter originating over a
wider (and mostly deeper) height range. \newch{Note that there is not 
always a nice time match between SJI 2796 and
SJI 1330/1400, sometimes the CBGs appear earlier, sometimes later in time.
An example of this mis-match between SJI 2796 and SJI 1330/1400
is shown in Figure~3 between 4400 s and 4600 s (see section 3.3). }

The light curves of the SJI CBGs and \cah\ as a function of time 
are shown in Figure~\ref{fig:sji1d}. In general 
CBGs show the largest variation in the 1400~\AA\ SJI filter, 
then in the 1330~\AA\ SJI filter, 
and the variability in the 2796~\AA\ SJI filter and \cah\ 
are the lowest. As mentioned above, there is a clear 
similarity between the SJI 1330 and 1400~\AA\ light curves and between 
SJI 2796~\AA\ and \newch{\cah. \cai\ matches} also nicely with the SJI 1330 and 
1400~\AA\ light curves (e.g., t=[8400,8600]s). 

Finally, but not least important, bright grains move
horizontally. This is seen
in Figure~\ref{fig:sjixt} as inclined trajectories of the grains with a
typical velocity of 7~km~s$^{-1}$ (see also Movie 1). 

\subsection{Origin of the CBGs emission}~\label{sec:emiss}

The SJI 1330~\AA\ filter is dominated by the \ion{C}{2}~1334 and
1336~\AA\ lines, the SJI 1400~\AA\ filter by the \ion{Si}{4}~1394 
and 1403~\AA, and the SJI 2796~\AA\ filter by the \mgka\ line. 
Is the CBG brightening in the SJI filters coming from these dominant
lines? We used rasters of long exposure observations taken with 
the IRIS spectrograph to address this question. 
Two different observations were used so that we can analyze all three IRIS
SJI filters (SJI 1330~\AA\ and 1400~\AA\ for 2014-Feb-15, SJI 2796~\AA\
and 1400~\AA\ for 2013-Oct-22). 

\begin{figure*}
  \includegraphics[width=0.99\textwidth]{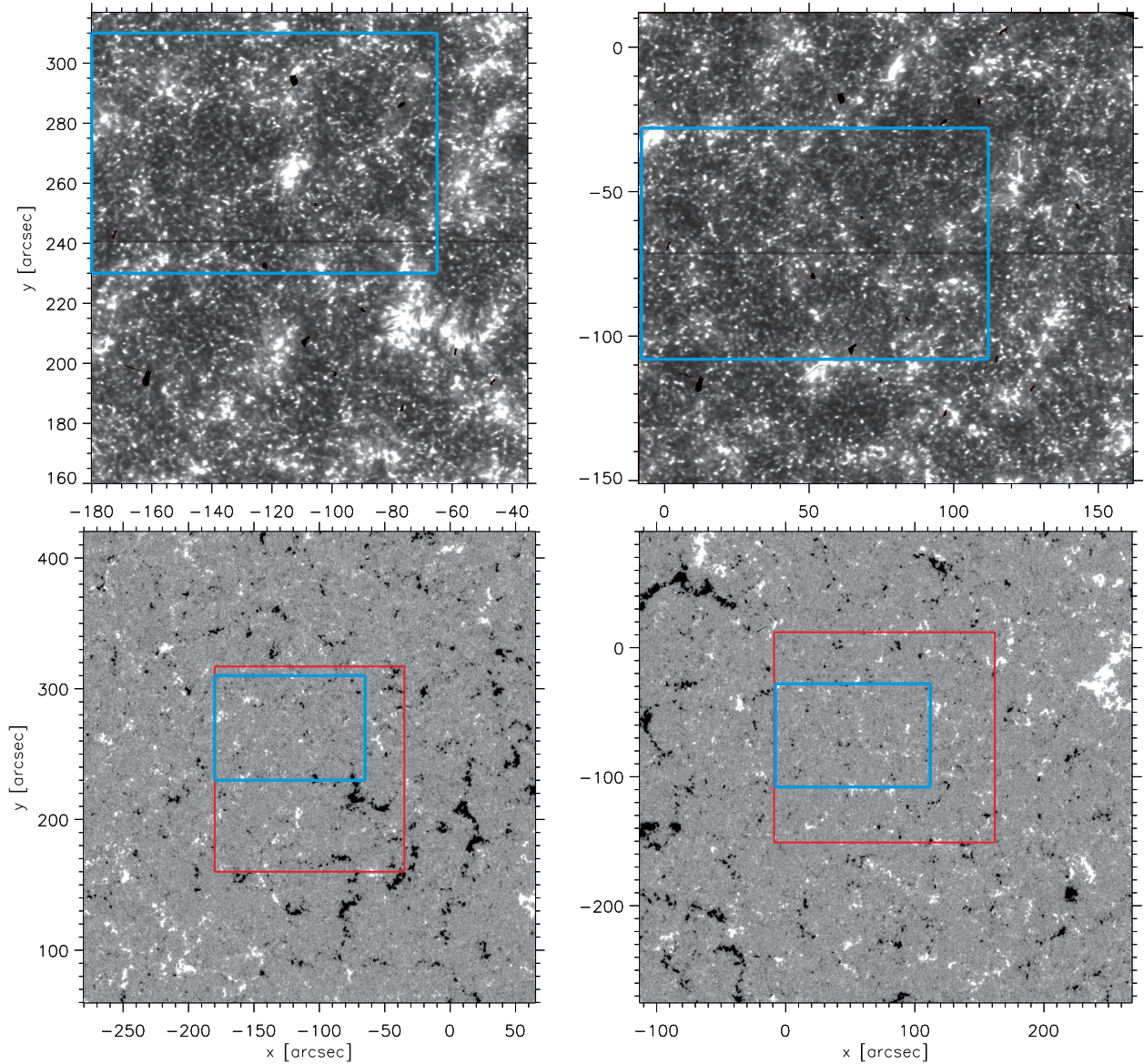}
 \caption{\label{fig:shmi} Context maps for the deep rasters shown in
   Figure~\ref{fig:intline}: SJI 1400 (top panels) and HMI  
   line-of-sight magnetograms (bottom panels). 
 The left panels are taken on 2013-October-22 at 11:30:30 UT and the
 right panels on 2014-February-25 at 20:50:31 UT.  
 The red box in the bottom panels delimits the FOV shown in the top panels and 
 the blue boxes mark the FOV shown in Figure~\ref{fig:intline}.
 The IRIS spectrograph slit is visible as the dark horizontal line in
 the middle of the SJI images (the satellite was rolled 90\degr\ with
 respect to solar North). }
\end{figure*}

For context, we show SJI 1400 images (top panels) 
and the line-of-sight magnetic field maps from HMI (bottom panels) in 
Figure~\ref{fig:shmi}. 
The context images serve to illustrate the low level of magnetic
activity and the ubiquitous presence of CBGs over the whole field of
view. 

\begin{figure*}
  \includegraphics[width=0.99\textwidth]{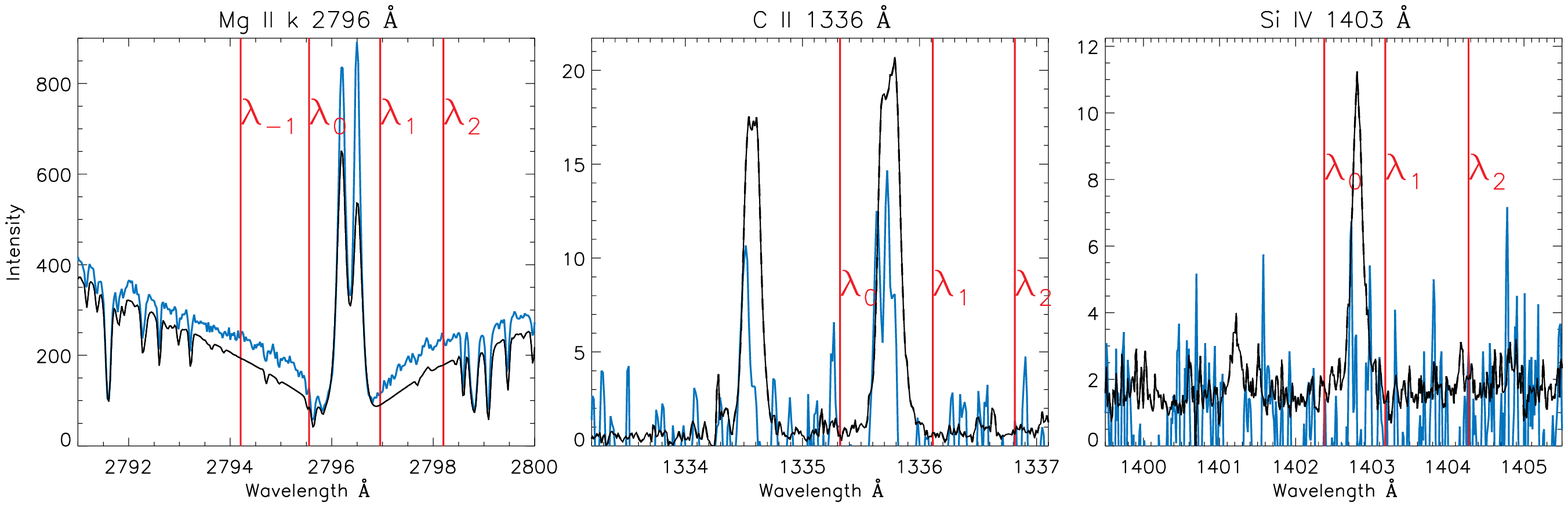}
 \caption{\label{fig:averplt} Observed spectral windows for the
   \ion{Mg}{2} k line (left), \ion{C}{2} lines (middle), and \si4\ (right). The black
   spectral profiles are spatially averaged over the full field of
   view of the IRIS raster of 2013 October 22 at 11:30:30 UT, including both internetwork and
   network regions. The blue spectral profiles are from one
   chromospheric bright grain (smoothed over 3 spectral pixels for
   \ion{C}{2} and \si4\ to suppress noise). The vertical red lines
   mark the spectral windows to construct the integrated line maps
   (between $\lambda_0$ and $\lambda_1$) and to estimate the 
   wing intensity for \mgk\ (between $\lambda_{-1}$ and $\lambda_0$, plus  
   between $\lambda_1$ and $\lambda_2$) and continuum intensity 
   for \ion{C}{2} and \si4\ (between $\lambda_1$ and $\lambda_2$).}
\end{figure*}

\begin{figure*}
  \includegraphics[width=0.99\textwidth]{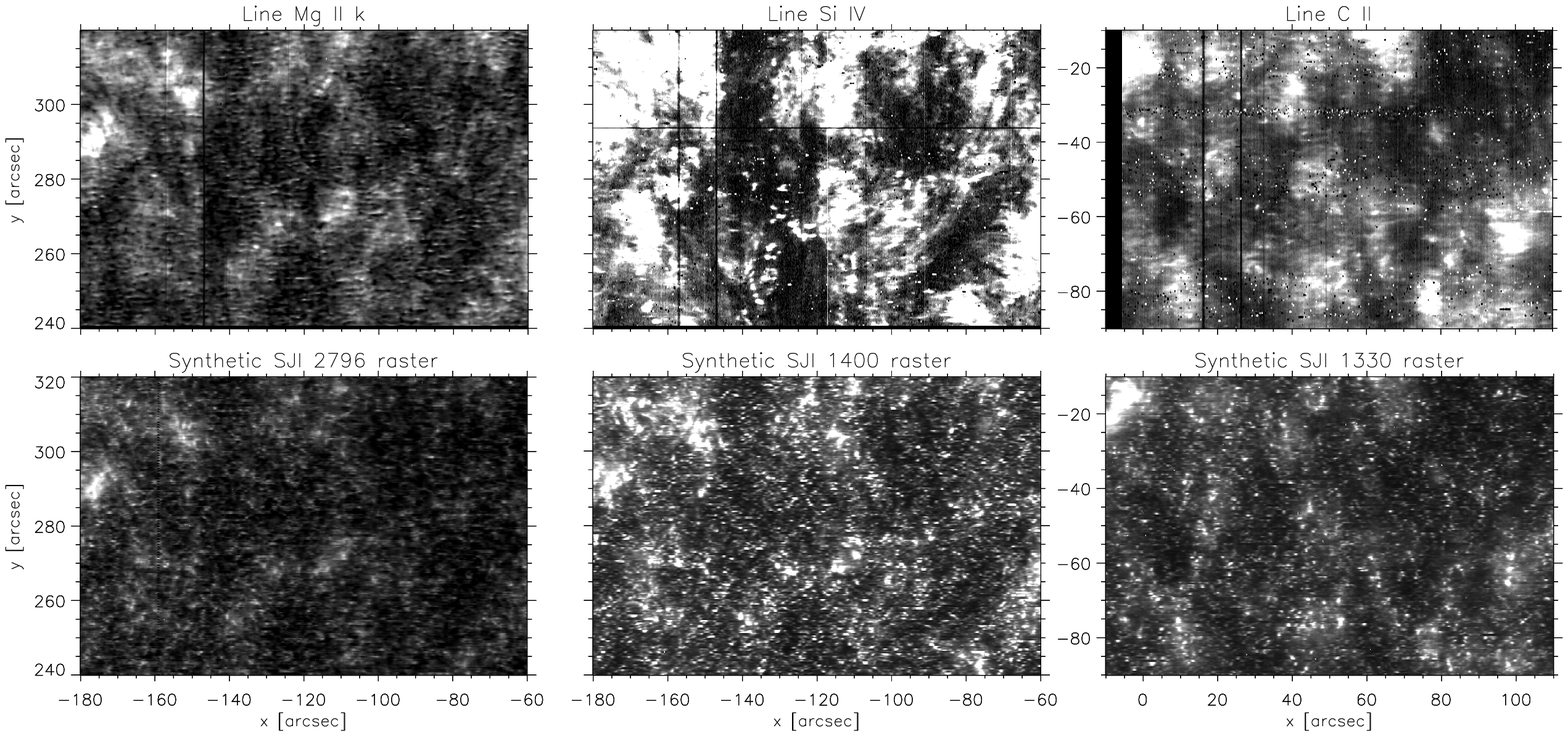}
 \caption{\label{fig:intline} Top row: Integrated intensity maps of the
   \mgk\ (left panel), \ion{Si}{4}~1403~\AA\ 
 (middle panel), and \ion{C}{2}~1336~\AA\ (right panel) lines from
 30~s exposure dense raster observations with the slit oriented align E-W. Bottom row: associated  
 \mgk\ (left panel), \ion{Si}{4}~1403~\AA\ (middle panel), and 
 \ion{C}{2}~1330~\AA\ (right panel) filter synthetic raster maps constructed from a series of SJI images 
 with time correction to match the raster. The field of view
 covers a very quiet internetwork region with only a few network
 elements near the edges (see context images in Figure~\ref{fig:shmi}). 
 The first two columns are from observations on 2013 October 22 at 11:30:30 UT,
 the right column is from observations on 2014 February 25 at 20:50:31 UT. The grey 
 scale ranges through the same normalized range for each channel and 
 is linear. See corresponding Movie~2.}
\end{figure*}

The 30~s exposure spectroheliograms allow us to search for possible line 
emission in \mgk, \ion{C}{2}, and \si4\ in CBGs in the IRIS SJI filter images. 
For this we integrated the spectral profiles as a function of wavelength 
and subtracted from \mgk\ the wings intensity and from \ion{C}{2} and \si4\ the 
continuum intensity in order to construct averaged line intensity maps 
(see the top panels of Figure~\ref{fig:intline}). 
The spectral windows used for integration are illustrated in Figure~\ref{fig:averplt}. 
The black profiles are averaged over the whole dataset which includes
network regions. 
Spectral profiles of one CBG are shown in blue (smoothed over three
spectral pixels for \ion{C}{2} and \si4\ to reduce noise). 
The \mgk\ wing intensity is calculated by averaging in wavelength from $\lambda_{-1}$ to $\lambda_0$ 
and from $\lambda_1$ to $\lambda_2$. The continuum intensity for the \ion{C}{2} and \si4\ 
lines is averaged from $\lambda_1$ and $\lambda_2$. The averaged continuum or \mgk\ wing intensities
are then subtracted from the average line profile intensity (between $\lambda_0$ and $\lambda_1$ ).
In Figure~\ref{fig:intline}, such integrated intensity maps of the dominant lines
from the raster are compared with the corresponding synthetic SJI raster maps (bottom panels). 

The raster maps were recorded with the slit oriented in the East-West
direction (along the $x$-axis in Figure~\ref{fig:intline}) and stepping in the
North-South direction (along the $y$-axis).
Horizontal bands with black and white pixels are caused by enhanced noise
from energetic particle hits on the detectors when the spacecraft
passed through the Earth's radiation belts (South Atlantic Anomaly). 
Vertical dark lines are due to artifacts on the slit. 
The corresponding synthetic SJI raster maps in the bottom row are corrected for the
temporal raster stepping: the maps are constructed from intensity
profiles neighboring the slit taken from SJI images that 
match the raster spectrograms. Note, we cannot get exactly 
co-spatial SJI data with the raster slit. 

The two \mgk\ maps (left panels in Figure~\ref{fig:intline} and Movie~2) are very
similar and are both dominated by CBGs and structures that appear
reminiscent of reversed granulation. SUNRISE observed also similar 
structures in \ion{Mg}{2}~k  \citep{Danilovic:2014fk}.
This shows that CBGs have a clear contribution from the \mgk\ line core.
Note that even though the SJI 2796 filter has rather narrow bandwidth
(4~\AA), there is still a significant contribution from the inner wings. 
For the construction of the integrated line intensity map in the top
panel, the wing contribution has been subtracted and there still is a 
significant CBG contribution.
The wing map (not shown) is also dominated by CBGs which leads to the
conclusion that CBG signals in the SJI 2796 filter comes both from the
\mgk\ line core and inner wings and explains why the contrast in both 
raster maps is not identical.

The \ion{C}{2} integrated line intensity map (top right panel of
Figure~\ref{fig:intline}) is rather different from the SJI 1330 map 
(bottom right) although close inspection reveals faint presence 
of some CBGs in the \ion{C}{2} map (see right panels on Movie 2). 
One clear example \newch{of the latter} is the selected case shown as the blue profile in  
Figure~\ref{fig:averplt} which is a CBG with a clear signal in \ion{C}{2}
and no signal in \si4. The \sia4\ map is filled with 
fibrils  \newch{\citep[middle top panel of 
Figure~\ref{fig:intline}, studied in detail by, e.g., ][]
{De-Pontieu:2007cr,Pereira:2014eu,Tian:2014fp}} and only in a few regions  
one can see weak grains such as around 
$(x,y)=(-140\arcsec,250\arcsec)$ and $(x,y)=(-120\arcsec,245\arcsec)$. 
The fibrilar structures are mostly saturated to visualize the faint CBGs. 
The brightness of the few grains in \si4\ and \ion{C}{2} intensities 
is lower than the fibrils; therefore, the CBG's 
\ion{Si}{4} and \ion{C}{2} emission 
is too weak to contribute significantly to the grain signal 
observed in SJI 1400~\AA\ 
and 1330~\AA\ (Figure~\ref{fig:intline} and Movie~2). 
If these spectral lines do not produce the CBG's emission in these two
filters, where does the signal come from?
Presumably, the emission of CBGs in the 1400~\AA\ and 1330~\AA\ SJI filters 
comes from the continuum. 
The observed continuum from the IRIS spectrograph is very weak and 
often does not exceed the digitization threshold, i.e.,  
we could not directly confirm if most of the CBGs emission in these two 
filters comes from the continuum by constructing integrated continuum 
maps like we did for the spectral lines.
The continuum windows in the IRIS FUV spectrograms are simply too
narrow compared to the 55~\AA\ bandpass of the SJI filters to
make a direct comparison.
Consequently, we postulate that grains in SJI 1400 and 1330~\AA\ 
are mostly caused by an integrated continuum over the wide filter
bandpass.

\begin{figure*}
  \includegraphics[width=0.99\textwidth]{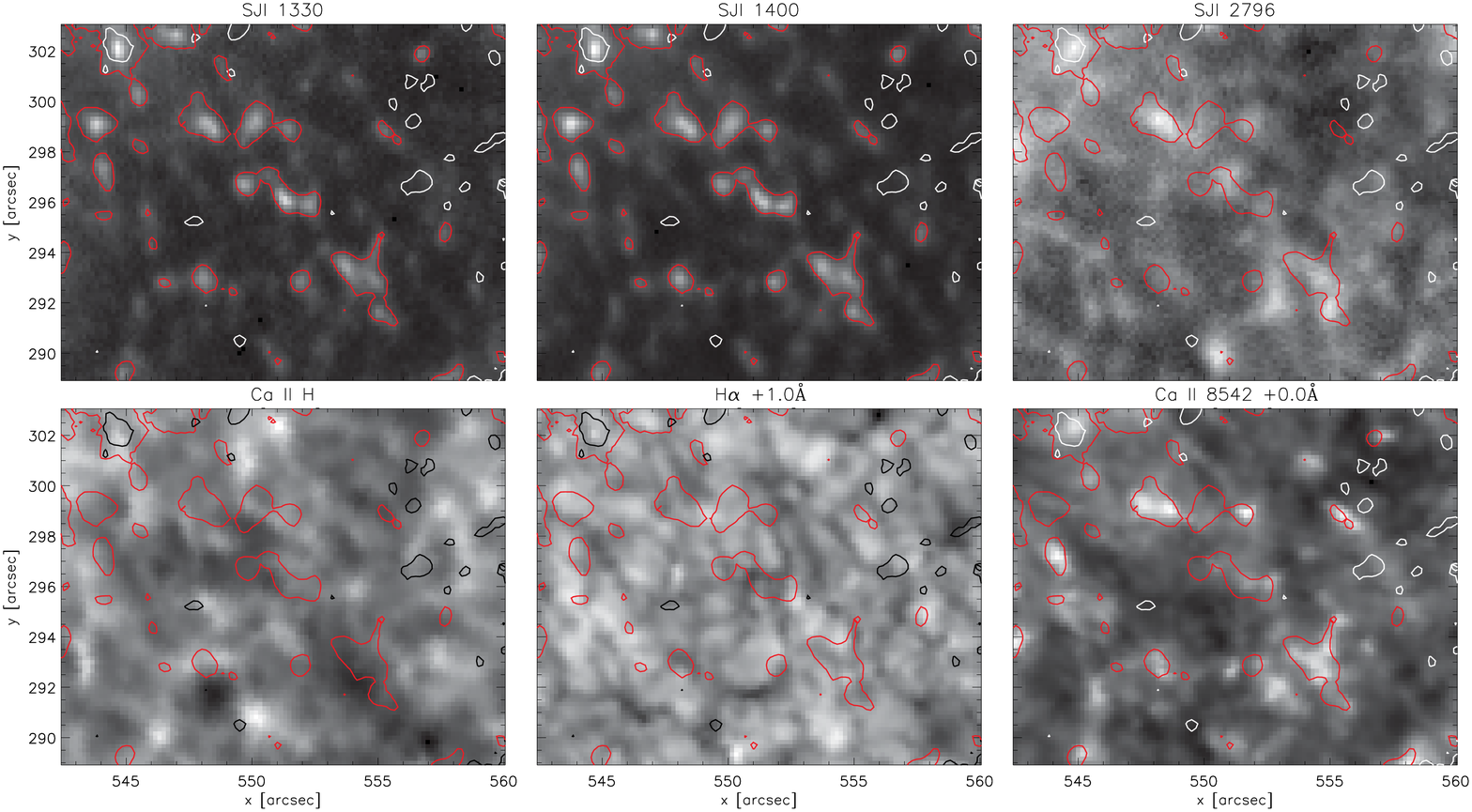}
 \caption{\label{fig:sjisst}  CBGs in UV (top row: SJI 1330, SJI
   1400, and SJI 2796) and optical (bottom row: line core of \cah, \Halpha\, 
   and \cai ) diagnostics. Red contours outline CBGs defined in SJI 1400, 
   black/white contours outline Stokes V signals. See corresponding Movie 3.
   The CBG locations match well, especially in the UV images and the Ca II 8542 line core.}
\end{figure*}

By comparison with the spectrally resolved data for \Halpha\ and \cai\ from
CRISP, we can make a rough estimate of the formation height of the SJI
1400~\AA\ and 1330~\AA\ bright grains. 
For this, we focus on CBGs within coronal holes and quiet sun 
internetwork, i.e., far from photospheric magnetic bright points. 
We use the \fe1\ Stokes V maps to select CBGs that are in the
internetwork: Figure~\ref{fig:sjisst} and Movie~3 show CBGs identified in SJI
1400 as red contours in different diagnostics. Black/white contours
outline regions with enhanced Stokes V signal.
In order to find the origin of the CBGs emission in SJI~1330~\AA\ (top left panel), 
1400~\AA\ (top middle panel) and 2796~\AA\ (top right panel) we compare them 
with \cah\ core (bottom left panel), \Halpha\ at various wavelength positions 
(bottom middle panel), and \cai\ at different wavelength positions (bottom right panel)
shown in Movie~3. 
Most of the emission seems to come from the network, but still, bright grains 
appear frequently all over in coronal holes and quiet sun
internetwork (see Figures~\ref{fig:sjixy} and~\ref{fig:shmi}). 
SJI bright grains tend to be located at the outer part of dark region 
boundaries in \cah , i.e., at the bright region of the reverse granulation. 
\Halpha\ maps differ the most from any of the SJI filters. Only in the line 
core of \Halpha\ can one see a few similarities. Figure~\ref{fig:sjisst} and Movie~2 
reveal a rather large overlap between \cai, \cah\ maps, and 1400~\AA\ and 
1330~\AA\  SJI filters. The best match of SJI 1330~\AA, and
1400~\AA\ CBGs is with bright grains as seen in the line core intensity of \cai\ 
(see Movie~3). Note that we have integrated the line core intensity
over  $\pm 0.1$~\AA\ with respect to the wavelength of the darkest core intensity
in order to avoid cross-talk between Doppler-shift and core intensity variations. The
correlation between \cai\ core intensity and IRIS slitjaws 
is also noticeable in the lightcurves shown in Figure~\ref{fig:sji1d}.  
\newch{The continuum of the SJI 1330 and 1400~\AA\  filters is normally
formed at a lower height than the core of the \cai\ line (roughly $0.8$~Mm 
\citet{Vernazza:1981yq} versus $1.3$~Mm
\citet{Cauzzi:2008jk}). However, in a dynamic atmosphere
the intensity of UV continua may get a large contribution
from the shock even when the shock is significantly higher than
the optical depth unity height
\citep{Carlsson:1995ai}. This explains the similarity between
the SJI 1330/1400~\AA\ and the \cai\ core intensities. }

\subsection{Spectral analysis}~\label{sec:dopp}

\begin{figure*}
  \includegraphics[width=0.99\textwidth]{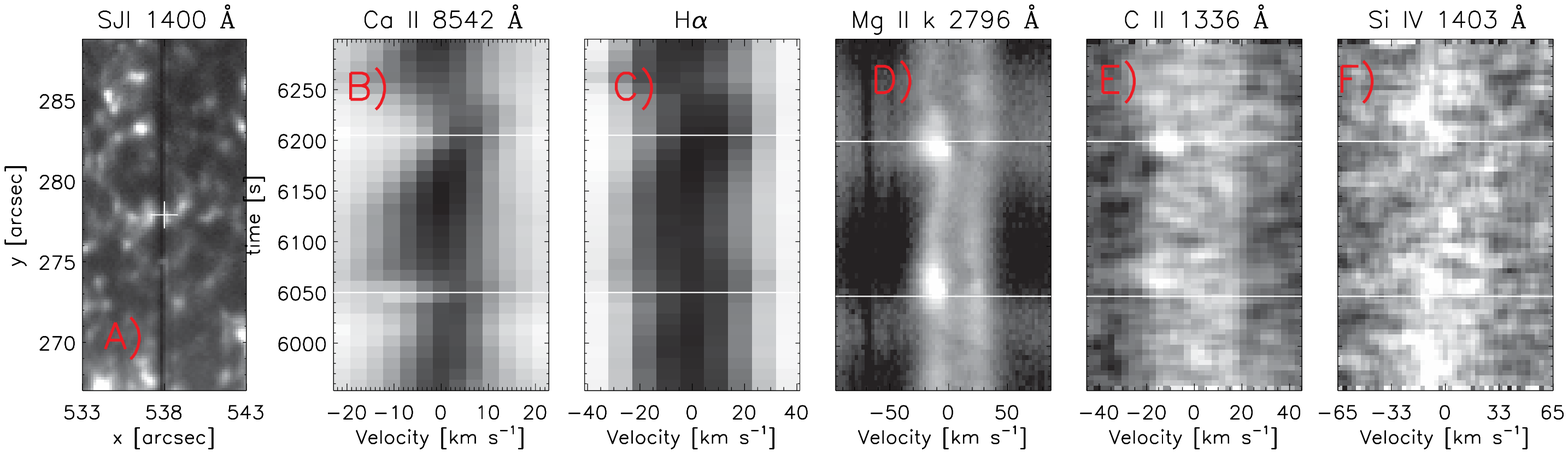}
 \caption{\label{fig:prof} Spectral profiles as a function of time for 
 \cai, \Halpha, \mgk, \ion{C}{2}, and \ion{Si}{4} are shown respectively in 
 panels B--F. The context image is shown in panel A at 
 time 6200~s (marked with the top horizontal white line in panels B-F) where the CBG is 
 captured by the slit (top white cross). This CBG shows also weak signal in
 \ion{C}{2} (panel E) and \si4\ (panel F). We binned
 the signal in \ion{C}{2} over 5 spatial pixels, two spectral pixels, and two consecutive 
 temporal snapshots (i.e., 20 spectra). For \si4, we binned over 7 spatial pixels, two spectral pixels 
 and two consecutive temporal snapshots (i.e., 28 spectra). \newch{The white horizontal 
lines are drawn to guide the eye for the time of the Ca II 8542 blue
wing brightening.}}
\end{figure*}

Both SST and IRIS provide spectral information as a function of  time for CBGs. 
Figure~\ref{fig:prof} shows \cai, \Halpha, \mgk, \ion{C}{2}, and \si4\   
(panels B-F) spectral profiles as a function of time for 
a number of SJI CBGs. The context SJI 1400~\AA\
image is shown in panel A at time=6200~s (marked with the upper horizontal white line in panels 
B--F).  
The $\lambda$-t plots of the chromospheric spectral lines reveal a wave pattern 
(panels B--D). Two nice examples of CBGs are at $t=6050,$ and $6200$~s in the 
figure (marked with the two horizontal white lines). 

At the beginning of the bright grain evolution, 
the core of the chromospheric lines are blue-shifted 
up to $-4$~km~s$^{-1}$ for \cai\ and \Halpha , and $-10$~km~s$^{-1}$ for 
\mgk\ for the absorption part of the spectral line. These Doppler velocities 
are roughly sonic, i.e., their amplitudes in Doppler velocity shift are rather 
small, most likely because the centroid velocity does not capture shocks 
since these lines are optically thick \newch{\citep{Carlsson:1997tg}}. Only \newch{at 
the moment of maximum Doppler shift}, the k2v peak 
shows a large intensity increase in both CBGs. 
It is known that a shock passing through the middle 
chromosphere produces an increase in density and temperature leading to a 
large intensity increase in the peak of \mgk 2v \citep[][detail the formation of 
the \mgh~\&~k lines in a simulated atmosphere]{Leenaarts:2013ij,
Leenaarts:2013hc,Pereira:2013ys} similar to the 
asymmetric intensity increase observed in the strong \ion{Ca}{2} lines (panel B)
\citep{Rutten:1991fk}. It is also interesting to see that at the later 
stage of the wave in \cai\ and \Halpha, the blue wing shows an asymmetry 
with an absorption contribution shifted up to $-30$~km~s$^{-1}$. 
As mentioned in the previous section, only some cases reveal some signal 
in the transition region lines, for instance the CBG example at $t=6200$~s, 
weak signals of \ion{C}{2} ($-15$~km~s$^{-1}$, panel E) and \ion{Si}{4} 
($-20$~km~s$^{-1}$, panel F) can be discerned. 
Similarly, the other example (at $t=6050$~s) has an increase in \ion{C}{2} intensity with comparable 
Doppler shifts.
The signal in \si4\ is very noisy and weak in CBGs (see for instance 
Figures~\ref{fig:averplt}, and~\ref{fig:intline}). 
Still, a weak \si4\ signal around $t=6200$s can be discerned at $-20$~km~s$^{-1}$.
Therefore, the Doppler shift, as expected for an upward traveling wave 
into lower density regions, increases with formation height 
from \cai\ ($\sim-4$~km~s$^{-1}$), \Halpha\ ($\sim-5$~km~s$^{-1}$) \mgk\ ($\sim-10$~km~s$^{-1}$), 
\ion{C}{2} ($\sim-15$~km~s$^{-1}$), to \si4\ ($\sim-20$~km~s$^{-1}$). However, one must be really 
careful with interpreting the shifts in the core of \cai, \Halpha, and \mgk\ as Doppler shifts since these 
are optically thick. 

Temporal evolution of \cai,  \Halpha, and \mgk\ spectral profiles  for a pixel in the 
internetwork in a coronal hole have a wave pattern with, sometimes, a time difference of a 
few seconds between each of them. For instance, the CBG at t=6200 s
shows the brightening in the blue wing in \cai\ earlier than the brightening in 
\ion{Mg}{2} k2v peak, whereas these two brightenings are reversed in time
of appearence in the CBG at t=6050 s.  Going through the data (not showing everything here) 
we find a similar number of appearances for both cases. 
If we for a moment make the naive assumption of the \cai\ brightening being formed at lower heights 
than \Halpha\ and even lower heights than \mgk\ one may interpret this as 
one case being an upward propagating wave and the other one as a downward propagating
wave. Far from this, Panel D) reveals that both CBGs come from upward propagating waves since
there first appears an increased intensity in the wings of \mgk\ propagating inwards towards the 
line core. This is a clear sign of an upward propagating wave. Once again, similar to the fact that one must be careful to interpret 
the line core shifts as Doppler shifts, caution must be exercised in the interpretation of the brightening
at the blue side of the line profile. \newch{This brightening comes about through an interplay between a local
source function maximum and the velocity gradient in optically thick line formation 
\citep[for a detailed discussion, see][]{Carlsson:1997tg}. Therefore, normal height of formation arguments must be treated with caution -
numerical simulations combined with non-LTE radiative transfer calculations giving synthetic observables 
may help with a more detailed interpretation}.

\section{Discussion and conclusions}~\label{sec:dis}

We analyzed bright grains observed in the internetwork of coronal holes 
and the quiet sun with the IRIS SJI 1330, 1400, and 2796~\AA\ filters. 
Chromospheric bright grains have a lifetime of roughly  1.5-2 minutes in the SJI 
1330, 1400~\AA\ filters and 2.5 minutes in the SJI 2796~\AA\ filter and \cah. 
\newch{The longer duration in the SJI
2796~\AA\ filter is, most likely, due to the fact that this spectral
passband better covers the atmospheric height range for which the
CBGs give enhanced emission. The filter (4~\AA\ passband) includes
both inner wings and line core of Mg k so that this channel covers
CBGs from the upper-photosphere to the chromosphere 
\citep{Pereira:2013ys}.  }
This explains why this channel shows CBGs for a longer period of time
than the other two SJI channels. We find a good correlation
between the light curves of the SJI 1330~\AA\ and 1400~\AA\ filters
and between the SJI 2796~\AA\ filter and \cah. 
\newch{It is clear that the CBG signal} through the  SJI 2796~\AA\ filter comes from the same 
range of heights as \cah. In addition, the intensity observed in the SJI 
1330~\AA\ and 1400~\AA\ filters must come from the same region. 
From a comparison of integrated line maps and SJI images, we conclude
that CBGs in these two IRIS filters are dominated by the continuum intensity and
not by the \ion{C}{2} and \si4\ emission lines in these filters.
Therefore, CBGs observed with the IRIS 1330 and 1400~\AA\ slit-jaws sample conditions between 
the upper photosphere and  middle chromosphere, as a matter of fact, they match 
rather nicely with the \cai\ line core taken with the Swedish 1-m Solar Telescope. 
Moreover, bright grains in SJI 1400~\AA\ and 1330~\AA\  
\newch{ appear towards the
end of the lifetime of the SJI} 2796~\AA\ features. 
Assuming that these events travel upwards,  the signal may 
come from roughly the middle chromosphere since 
the grains in SJI 1400 and 1330~\AA\  seem to form more or less 
at similar heights as the core of \mgk.

If grains are a consequence of shocks and the 
density drops drastically between the chromosphere and transition region, 
one may expect a greater amplitude of the shock in the transition region. 
Note that most of the CBGs signal in the SJI filters comes from the 
chromosphere and not from the transition region. 
Using the IRIS spectrograph, we were able to reveal that CBGs sometimes %?
have weak and smooth emission in 
\ion{C}{2} \citep[in agreement with][]{Carlsson:1997ys} and in a few 
cases in \si4\ too, i.e., such waves can reach the transition region. 
Their emission get blurry and extremely weak, most 
likely due to expansion of the shock when it travels through the chromosphere
and reaches overlying loops. Further studies of CBGs simulations 
should seek if overlaying loops and canopy structures can lead to weak 
emission in transition region lines and/or stop the propagation  
of the shocks into the transition region. Another open question is if shocks with 
weak signals in the transition region can get into the corona.
 
For the first time we tie the same shock in \cai, \Halpha, and \mgk\, and 
with a weak signal in \ion{C}{2}, and in a few cases in \si4. In addition, 
we match these signatures in the various spectral profiles 
with CBGs in the IRIS SJI filters. 
\cai, \Halpha, and \mgk\ profiles reveal shock patterns with amplitudes of 
roughly $4$~km~s$^{-1}$ for \cai\ and \Halpha\ and $10$~km~s$^{-1}$ for \mgk\ and 
their lifetime is roughly $\sim1.5$~minutes. 
\ion{C}{2} and \si4 revealed only a weak brightening around $20$~km~s$^{-1}$. 
However, one must be careful to interpret the line core shifts of the chromospheric lines
because they are optically thick. As a matter of fact, we see that 
the \cai\ brightening is sometimes produced before and other times after the 
intensity increase of \ion{Mg}{2}~k2v. However, in both cases \mgk\ reveals
that the wave is propagating upwards since the enhancement in intensity appears
first in the wings and propagates in time towards the core of the line. 

\section{Acknowledgments}

We gratefully acknowledge support by NASA grants NNX11AN98G,
NNM12AB40P and NASA contracts NNM07AA01C (Hinode), and NNG09FA40C 
(IRIS). This work has benefited from discussions at 
the International Space Science Institute (ISSI) meeting on 
``Heating of the magnetized chromosphere'' from 24-28 February, 2014, 
where many aspects of this paper were discussed with other colleagues.
To analyze the data we have used IDL and the Solar SoftWare library (SSW).
The Swedish 1-m Solar Telescope is operated by the Institute for Solar Physics 
of the Royal Swedish Academy of Sciences in the Spanish Observatorio del 
Roque de los Muchachos of the Instituto de Astrof\'{\i}sica de Canarias.  
This research was supported by the Research Council of Norway and by the 
European Research Council under the European Union's Seventh Framework 
Programme (FP7/2007-2013) / ERC Grant agreement nr. 291058. Thanks are 
also due to Hakon Skogsrud, Eamon Scullion, and Ada Ortiz for doing the 
observations at the SST. 

\bibliographystyle{aa}

\end{document}